\documentstyle[prl,aps,twocolumn,epsfig]{revtex}

\begin{document}
\topmargin -40pt
\draft
\twocolumn[
\hsize\textwidth\columnwidth\hsize\csname @twocolumnfalse\endcsname

\title{Enhancement of tunneling from a correlated 2D electron 
system by a many-electron M\"ossbauer-type recoil in a magnetic field}
\author{ M.I.~Dykman$^{(a)}$, T. Sharpee$^{(a)}$, and P.M.~Platzman$^{(b)}$}
\address{$^{(a)}$Department of Physics and Astronomy, 
Michigan State University, East Lansing, Michigan 48824\\ $^{(b)}$Bell
Laboratories, Lucent Technologies, Murray Hill, New Jersey 07974}
\date{\today} 

\maketitle
\begin{quote} 
We consider the effect of electron correlations on tunneling from a 2D
electron layer in a magnetic field parallel to the layer. A tunneling
electron can exchange its momentum with other electrons, which leads
to an exponential increase of the tunneling rate compared to the
single-electron approximation. The effect depends on the interrelation
between the {\it dynamics} of tunneling and momentum exchange. The results
explain and provide a no parameter fit to the data on electrons on
helium. We also discuss tunneling in semiconductor heterostructures.

\end{quote} 
] 
\narrowtext 

Low density two-dimensional electron systems (2DES) in semiconductor
heterostructures and on liquid helium are among the most ideal
many-electron systems. Such systems display strong effects of the
electron-electron interaction, including those specifically related to
electron correlations \cite{Abrahams-00,Andrei_book}. They
show up dramatically in various unusual transport properties. 
One of the most broadly used techniques for investigating
many-electron effects is tunneling
\cite{DasSarma-97}, a recent example being the observation
\cite{Spielman-00} of the giant increase of interlayer tunneling in
double-layer heterostructures, apparently related to the onset of
interlayer correlations.

For electrons on helium, an exponentially strong deviation from the
single-electron rate of tunneling transverse to a magnetic field has
been known experimentally since 1993 \cite{Andrei-93}, but remained
unexplained.  Such a field couples the
tunneling motion away from the 2DES to the in-plane degrees of
freedom. The effect of the field and the role of electron correlations
cannot be described by a simple phenomenological tunneling
Hamiltonian.

In this paper we provide a theory of tunneling from a correlated 2DES
in a magnetic field ${\bf B}$ parallel to the electron layer. We show,
using the model of a Wigner crystal (WC), that the tunneling is
affected by the interelectron momentum exchange and its {\it
dynamics}, which
is largely determined by short-range order. We discuss tunneling from
2DES on helium and in single quantum well heterostructures. The
results explain and give a no parameter fit to the experimental data
\cite{Andrei-93}, see Fig.~1. They suggest new types of experiments
which involve tunneling through broad barriers and will be sensitive
to short-range order in a 2DES.

Electron correlations change the tunneling rate by effectively
decreasing the single-electron magnetic barrier. This barrier emerges
because, when an electron tunnels from the layer (in the
$z$-direction), it acquires an in-plane Hall velocity $v_H=\omega_c z$
in the ${\bf B} \times \hat{\bf z}$ direction and the corresponding
in-plane kinetic energy $m\omega_c^2z^2/2$, where $\omega_c=eB/mc$ is
the cyclotron frequency. Respectively, the energy for motion along the
$z$-axis is decreased, or the tunneling barrier is increased by
$m\omega_c^2z^2/2$.

In a correlated 2DES, the tunneling electron exchanges its Hall
momentum with other electrons, thus decreasing the energy loss
\cite{Shklovskii}. This is somewhat similar to the M\"ossbauer effect
where the momentum of a gamma quantum is given to the crystal as a
whole \cite{Levitov-96}. In our case, the effect is very sensitive to the electron
dynamics. If the rate of the interelectron momentum exchange
$\omega_p$ exceeds the reciprocal duration of underbarrier motion in
imaginary time $\tau_f^{-1}$, then in-plane velocities of all
electrons are nearly the same, and the Hall velocity is $v_H\propto 1/N
\to 0$ ($N$ is the number of electrons).  In this adiabatic limit 
the effect of the magnetic field on tunneling is fully
compensated. For $\omega_p\tau_f\sim 1$ a part of the tunneling
energy goes to WC phonons, yet the $B$-induced suppression of
tunneling is largely reduced.

\begin{figure}
\begin{center}
\epsfxsize=2.6in                
\leavevmode\epsfbox{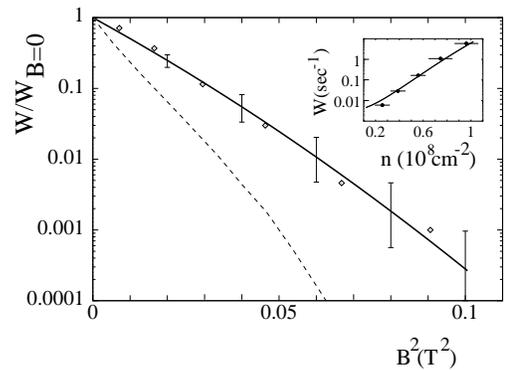}
\end{center}
\caption{The rate of electron tunneling  from helium surface $W(B)$  
as a function of the magnetic field $B$ for the electron density
$n=0.8\times 10^8 {\rm cm}^{-2}$ and the calculated pulling field ${\cal
E}_{\perp}=24.7$~V/cm (solid curve). Lozenges show the experimental data
\protect\cite{Andrei-93}. The error bars correspond to the 
uncertainty of the experimental parameters.  The dotted curve is the
calculation \protect\cite{Andrei-93} for $\rm T=0.04K$ without
inter-electron momentum exchange.  Inset: comparison of the present
theory for $B=0$ with the experimentally measured density dependence
of the tunneling rate.}
\label{fig:exponent}
\end{figure}

In a strongly correlated system, where the electron wave functions
overlap only weakly, one can ``identify'' the tunneling electron.  Its
out-of-plane motion for $B=0$ is described by the Hamiltonian
\begin{equation}
\label{H0}
H_0={p_z^2 \over {2 m}}+U(z).
\end{equation}
The potential $U(z)$ has a well in which the electron occupies the
ground state, with energy $E_g$. The well is separated by a tunneling
barrier from extended states with a quasicontinuous spectrum,
cf. Fig.~2 below. We assume that the tunneling length $L$ is much less
than the average inter-electron distance $\sim n^{-1/2}$, where $n$ is
the electron density. Then small-amplitude in-plane electron
vibrations about lattice sites are only weakly coupled to tunneling
for $B=0$
\cite{AP-90}. We neglect this coupling.

A magnetic field parallel to the electron layer mixes up
the in-plane and out-of-plane motions. The Hamiltonian of the
tunneling electron and phonons of the WC  is $H=H_0 +H_v+H_B$ with
\begin{equation}
\label{Hv}
H_v= {1\over 2}\sum_{{\bf k},j}\left[m^{-1}{\bf p}_{{\bf k}j}{\bf
p}_{-{\bf k}j} + m \omega_{{\bf k}j}^2 {\bf u}_{{\bf k}j}{\bf
u}_{-{\bf k}j}\right]
\end{equation}
and
\begin{equation}
\label{interaction}
H_B={1 \over 2} m\omega_c^2 z^2 - \omega_c z N^{-1/2}
\sum_{{\bf k},j}[\hat{\bf B}\times{\bf p}_{{\bf k}j}]_z.
\end{equation}
Here, ${\bf p}_{{\bf k}j}$, ${\bf u}_{{\bf k}j}$, and $\omega_{{\bf
k}j}$ are the momenta, displacements, and frequencies of the normal
modes of the 2D Wigner crystal with the wave vector ${\bf k}$,
respectively ($j=1,2$). We assumed that the equilibrium in-plane
position of the tunneling electron is at the origin. Then its in-plane
momentum is ${\bf p} =N^{-1/2}\sum{\bf p}_{{\bf k}j}$.

The Hamiltonian $H_B$ couples the out-of-plane motion to lattice
vibrations. The problem of many-electron tunneling is thus mapped onto
a familiar problem of a particle coupled to a bath of harmonic
oscillators \cite{Feynman,Leggett}, with the coupling strength
controlled by the magnetic field. However, there are two distinctions
from the standard formulation. First, the coupling mixes together the
particle {\it coordinate} $z$ and the {\it momenta} of the lattice.
These two quantities have different symmetry with respect to time
inversion. Because of broken time-reversal symmetry, the general
problem of tunneling in a 3D potential in a magnetic field requires a
special approach, which was developed earlier for an isolated particle
\cite{us}. For the present model, the problem is simplified by the fact that
in-plane motion is harmonic vibrations and the coupling is independent
of ${\bf u}_{{\bf k}j}$ \cite{Leggett}.

The second distinction arises, because for 2DES the potential well
$U(z)$ is strongly nonparabolic near the minimum (cf. Fig.~2). As a
result, the standard instanton technique \cite{Langer} does not apply
\cite{Ao}.

We will evaluate the tunneling rate in the WKB approximation.  In the
presence of a magnetic field it is convenient to look for the WKB wave
function under and behind the barrier in the momentum representation
with respect to phonon variables,
\begin{equation}
\label{psi}
\psi(z,
\{{\bf p}_{{\bf k}j}\}) =\exp[iS(z,\{{\bf p}_{{\bf k}j}\})],\quad \hbar =1,
\end{equation}
and make a canonical transformation so that ${\bf p}_{{\bf k}j}$ and
 $-{\bf u}_{{\bf k}j}$ be new canonical coordinates and momenta.

To the lowest order in $\hbar$, the action $S$ in (\ref{psi}) can be
obtained from the Hamiltonian equations for the trajectories of the
system,

\begin{eqnarray}
\label{eom}
\dot S = p_z\dot z& -& \sum_{{\bf k}j} {\bf u}_{{\bf k}j}\dot{\bf
p}_{{\bf k}j},\quad 
\dot z = {\partial H \over \partial p_z},\quad \dot p_z
=-{\partial H \over \partial z}\nonumber\\
&&\dot {\bf u}_{{\bf k}j}={\partial H \over \partial
{\bf p}_{{\bf k}j}},\quad
\dot{\bf p}_{{\bf k}j}=-{\partial H \over \partial {\bf u}_{{\bf k}j}}.
\end{eqnarray}
In the $(z, \{{\bf p}_{{\bf k}j}\})$-representation, the Hamiltonian
equations (\ref{eom}) have time-reversal symmetry. This allows us to
solve them under the barrier in a standard way \cite{Leggett} by
keeping the coordinates $z, {\bf p}_{{\bf k}j}$ real and making the
momenta $p_z, -{\bf u}_{{\bf k}j}$, time $t=-i\tau$, and action
$S(z,\{{\bf p}_{{\bf k}j}\})=iS_E(z,\{{\bf p}_{{\bf k}j}\})$ purely
imaginary.

The Euclidean action $S_E(\tau)$ as a function of time is evaluated
along a multidimensional trajectory (\ref{eom}) that goes under the
barrier from the potential well to the boundary of the region which is
classically allowed to both the tunneling electron and the WC
vibrations. At this boundary one has to match the underbarrier
solution (with imaginary momenta) with the WKB solution behind the
barrier (with real momenta), and therefore
\begin{equation}
\label{bound_euclid}
p_z(\tau_f)=0,\; {\bf u}_{{\bf k}j}(\tau_f)={\bf 0},
\end{equation}
where $\tau_f$ is the imaginary time at which the boundary is reached.

We now discuss the initial conditions for the trajectories (\ref{eom}).
Typically, the characteristic intrawell localization length $1/\gamma$
in the potential $U(z)$ is small compared to the tunneling length
$L$. For large $\gamma L \gg 1$, the magnetic field may have strong
cumulative effect on the tunneling rate, even where it only weakly perturbs the
intrawell motion. Inside the well and close to it the electron
in-plane and out-of-plane motions are then separated. We can set
initial conditions at an arbitrary plane $z=z_0$ close to the well,
yet deep enough under the barrier so that the wave function
$\psi(z,\{{\bf p}_{{\bf k}j}\})$ is semiclassical. For a harmonic WC,
the dependence of $\psi$ on ${\bf p}_{{\bf k}j}$ is Gaussian. Then
from (\ref{psi})
\begin{eqnarray}
\label{initialS}
&&S_E(0)= \sum\nolimits_{{\bf k}j} {\bf p}_{{\bf k}j}(0){\bf p}_{{\bf
-k}j}(0)/ 2m
\omega_{{\bf k}j}\nonumber\\
&&{\bf u}_{{\bf k}j}(0)=-i{\bf p}_{-{\bf k}j}(0)/ m\omega_{{\bf k}j}.
\end{eqnarray}

In the cases of interest, the dependence of $\psi$ on $z$ is
exponential near the well, $\psi\propto \exp(-\gamma z)$. Therefore

\begin{equation}
\label{initial1}
z(0) = z_0,\quad p_z(0)= i\gamma = i\sqrt{2m[U(z_0)-E_g]}.
\end{equation} 
Under the barrier, the potential $U(z)$ varies on the scale bigger
than $1/\gamma$, and then $\gamma$ in (\ref{initial1}) is independent of
the exact position of the plane $z=z_0$.

Solving the linear equations of motion (\ref{eom}) for the phonon
variables ${\bf u}_{{\bf k}j}, {\bf p}_{{\bf k}j}$ with the boundary
conditions (\ref{bound_euclid}), (\ref{initialS}), we can eliminate
them, cf. \cite{Feynman}. Then $ S_E$ takes the form of a retarded
action for 1D motion,
\begin{eqnarray}
\label{tun_exp}
&S_E[z]=&{1\over 2}\int_0^{2\tau_f}d\tau_1 \left[ {m\over
2}\left({dz/ d\tau}\right)^2+U(z)-E_g\right.\nonumber\\
&& +{1\over 2}m\omega_c^2 z^2
(\tau_1)-\left(m \omega_c^2 / 4N\right)\sum_{{\bf
k}j}
\omega_{{\bf k}j} \nonumber\\
&&\left.\times\int_0^{\tau_1} d
\tau_2 z(\tau_1) z(\tau_2) \exp [{-\omega_{{\bf k }j}(\tau_1-
\tau_2)}]\right].
\end{eqnarray}
In (\ref{tun_exp}) we symmetrically continued the trajectory $z(\tau)$
from $\tau_f$ to $2\tau_f$, with $z(\tau_f+ x)= z(\tau_f-x)$ for
$0\leq x \leq \tau_f$, and set $z_0=0$.  The added section of the
trajectory corresponds to underbarrier motion from the boundary of the
classically accessible range back to the potential well. The tunneling
rate $W\propto \exp[-R]$, with $R=2\,{\rm min}S_E$.

For small magnetic fields, the field-induced correction to the
tunneling exponent (\ref{tun_exp})  is quadratic in $\omega_c$. It can be
calculated along the zero-field trajectory $dz/d\tau =
[2U(z)/m]^{1/2}$. This correction is always positive: magnetic field
decreases the tunneling rate. However, the correction is smaller than
in the absence of the electron-electron interaction.

Remarkably, although a part of the energy of the tunneling electron
goes to WC phonons, the tunneling rate increases with the increasing
phonon frequencies.  If the characteristic $\omega_{{\bf k}j}$ largely
exceed the reciprocal tunneling time $1/\tau_f$, then
$z(\tau_2)\approx z(\tau_1)$ in the second term in (\ref{tun_exp}). As
a result, the $B$-dependent terms in (\ref{tun_exp}) cancel each
other, and tunneling is not affected by the magnetic field at
all. This happens because, as the tunneling electron moves under the
barrier, its in-plane momentum is adiabatically transferred to the
entire WC, similar to the M\"ossbauer effect. This can be contrasted
with the case of an electron confined only inside the well but not
under the barrier. Here the magnetic barrier is reduced by a factor of
two compared to the free-electron case, but does not disappear
\cite{Shklovskii}.

We now apply the results to {\bf electrons on helium} and compare them
with the experiment \cite{Andrei-93}. We will use the Einstein model
of the WC in which all phonons have the same frequency $\omega_p$,
which we set equal to the characteristic plasma frequency $(2\pi
e^2n^{3/2}/m)^{1/2}$. The numerical results change only slightly when
this frequency is varied within reasonable limits, e.g., is replaced
by the root mean square frequency of the WC $\bar\omega$ equal to
\cite{Mark}
\begin{equation}
\label{omega_0}
\bar\omega=\left[\sum\nolimits_{{\bf k}j}
\omega^2_{{\bf k}j}/2N\right]^{1/2}\approx \left(4.45 e^2n^{3/2}/ m
\right)^{1/2}.
\end{equation}

For an electron which is pulled away from the helium
surface by the field ${\cal E}_{\perp}$, the potential $U(z)$ has the form

\begin{equation}
\label{U_on_helium}
U(z) = -\Lambda z^{-1} - |e{\cal E}_{\perp}|z -m\bar\omega^2z^2
\end{equation}
for $z>0$ (outside the helium). On the helium surface (located at
$z=0$), $U(z)$ has a high barrier $\sim 1$~eV which prevents the
electron from penetrating into the helium.

In (\ref{U_on_helium}), the term $\propto \Lambda = e^2(\epsilon-1)
/4(\epsilon+1)$ describes the image potential, $\epsilon\approx 1.057$
is the dielectric constant. The field ${\cal E}_{\perp}$ is determined
by the helium cell geometry and depends on the applied voltage and the
electron density $n$, cf.~\cite{Eperp}. The term $\propto
m\bar\omega^2\equiv e^2\sum^{\prime}|{\bf R}_l|^{-3}/2$ describes the
Coulomb field created by other electrons at their lattice sites ${\bf
R}_l$ (the ``correlation hole''
\cite{AP-90,Iye-80}), for  the
tunneling length $L< n^{-1/2}$. The conditions $1/\gamma\ll L\ll
n^{-1/2}$ are typically very well satisfied in the experiment, with
$1/\gamma = 1/\Lambda m \approx 0.7\times 10^{-6}$~cm, $L=
\gamma^2/2m|e{\cal E}_{\perp}|\sim 10^{-5}$~cm for typical ${\cal
E}_{\perp}\sim 10$V/cm, and $n^{-1/2} \sim 10^{-4}$~cm.

The magnetic field dependence of the tunneling rate calculated from
Eqs.~(\ref{eom}) - (\ref{initial1}) is shown in Fig.~1. The actual
calculation is largely simplified by the fact that, deep under the
barrier, the image potential $-\Lambda/z$ in (\ref{U_on_helium}) can
be neglected. The equations of motion (\ref{eom}) become then linear,
and the tunneling exponent $R=2S_E(\tau_f)$ can be obtained in
an explicit (although somewhat cumbersome) form, which was used in
Fig.~1. The correction to $R$ from the image potential is $\sim
1/\gamma L$. When this and other corrections $\sim 1/\gamma L$ are
taken into account, the theoretical curve in Fig.~1 slightly shifts
down (by $\alt 20\%$ even for strong $B$), which is {\it much less}
than the uncertainty in $R$ due to the uncertainties in $n$ and
${\cal E}_{\perp}$ in the experiment
\cite{Andrei-93}.  The theory is in excellent 
agreement with the experiment, with no adjustable parameters.

The dependence of the potential $U(z)$ on $n$ gives rise to the
density dependence of the escape rate $W(B)$ even for $B=0$. We
calculated the exponent and the prefactor in $W(0)$ by matching the
WKB wave function under the barrier for $1/\gamma\ll z\ll L$ with the
intrawell solution. The latter was sought in the form
$\psi(z)=z\exp[-A(z)]$. The function $dA/dz$ satisfies a Riccati
equation which can be solved near the well ($z\ll L$) by considering
the last two terms in (\ref{U_on_helium}) as a perturbation. When
calculated to the first order in this perturbation, $A$ allows to find
not only the exponent, but also the leading term in the prefactor in
the WKB wave function. The resulting tunneling rate is shown in the
inset in Fig.~1. It fully agrees with the experiment \cite{Goodkind}.

For {\bf semiconductor heterostructures}, tunneling in correlated
systems has been investigated mostly for the magnetic field $\bf B$
perpendicular or nearly perpendicular to the electron layer,
cf. \cite{Spielman-00}. The data on tunneling in a field parallel to
the layer refer to high density 2DESs \cite{Eisenstein}, where
correlation effects are small. We expect that tunneling experiments on
low-density 2DESs in parallel fields will reveal electron correlations
not imposed by the magnetic field, give insight into electron
dynamics, and possibly even reveal a transition from an electron fluid
to a pinned Wigner crystal with decreasing $n$.

The effect of a parallel magnetic field is most pronounced in systems
with shallow and broad barriers $U(z)$. For example, in a GaAlAs
structure with a square barrier of width $L=0.1\, \mu$m and height
$\gamma^2/2m=0.02$~eV, for the electron density $n= 1.5\times 10^{10}$
cm$^{-2}$ and $B= 1.2$~T we have $\omega_p\tau_0 \approx 0.6$ and
$\omega_c\tau_0 \approx 1$ ($\tau_0=mL/\gamma$ is the tunneling
duration for $n=B= 0$).
\begin{figure}
\begin{center}
\epsfxsize=2.8in                
\leavevmode\epsfbox{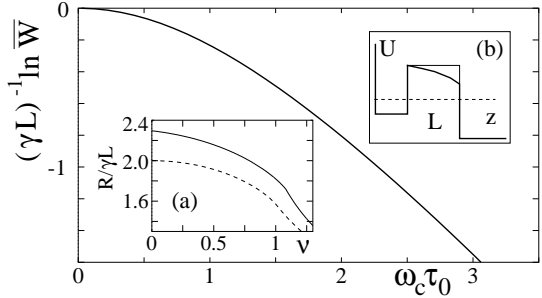}
\end{center}
\caption{Relative rate of tunneling $\bar{W} = W(B)/W(0)$ 
vs magnetic field for a 2D WC in a semiconductor heterostructure, with
$\bar\omega\tau_0=0.5$.  Inset (a): the tunneling exponent $R$ vs $\nu =
\sqrt{2}\bar\omega\tau_0$ for $\omega_c\tau_0=1.0$ (solid line) and $B=0$
(dashed line). Inset (b): the tunneling potential with (bold line) and
without (thin line) barrier reduction due to static electron correlations.}
\label{fig:semicond}
\end{figure}
\vspace{-0.1in}

Electron correlations give rise to a coordinate-dependent lowering of
the barrier, see Fig.~2. For $nL^2\ll 1$, $U(z)=\gamma^2/2m
-m\bar\omega^2z^2$, $0<z<L$ [we count $U$ off from the intrawell energy
level $E_g$]. The picture of tunneling depends on the parameter $\nu =
\sqrt{2}\bar\omega \tau_0 $. For
$\nu < 1$ the electron comes out from the barrier at the point $z=L$
where $U(z)$ is discontinuous, cf. Fig.~2b. In this most important
case, the boundary conditions (\ref{bound_euclid}) for the tunneling
trajectory should be changed to
\begin{equation}
\label{initial2}
z(\tau_f)=L, \quad {\bf u}_{{\bf k}j}(\tau_f)={\bf 0},
\end{equation}
but the tunneling exponent is still given by Eq.~ (\ref{tun_exp}).

For $B=0$ the tunneling exponent $R$ decreases with $n$, $R = \gamma
L[\nu^{-1}\arcsin\nu + (1-\nu^{2})^{1/2}]$ for $\nu < 1$, and $R=\pi
\gamma L/2\nu$, for $\nu >1$. Magnetic field causes $ R$ to increase
and the tunneling rate to decrease.  The effect is reduced by the
inter-electron momentum exchange.  The results for the Einstein model
of the WC with $\omega_{{\bf k}j}=\omega_p$ are shown in Fig.~2. The
inset of Fig.~2 shows how $R$ is decreased by the electron
correlations even for $B=0$.

We have used the model of a WC to analyze the effect of electron
correlations on tunneling in a magnetic field parallel to the electron
layer. We showed that the electron-electron interaction gives rise to
an exponential increase of the tunneling rate compared to its
single-electron value in a strong magnetic field. The effect is
determined by the interrelation between the frequencies of in-plane
electron vibrations and the reciprocal tunneling time. For long
tunneling time, the physics of large changes in the decay rate is
closely tied to the physics of the recoilless fraction in the
M\"ossbauer effect.  Since the major contribution comes from the
short-wavelength vibrations, the results should apply not only to WCs,
but also to all 2DESs with short-range order. Our results
give a quantitative no-parameter fit to the experimental data
\cite{Andrei-93} on tunneling of strongly correlated electrons
on helium.

This research was supported in part by the NSF through Grant No. PHY-0071059.


\begin{thebibliography}{99} 

\vspace*{-0.4in}
\bibitem{Abrahams-00} E. Abrahams, S.V. Kravchenko, and M.P.~Sarachik,
cond-mat/0006055.

\bibitem{Andrei_book} {\it Two-Dimensional Electron Systems on Helium and
other Cryogenic Substrates}, ed. by E. Andrei (Kluwer, NY 1997).

\bibitem{DasSarma-97} {\it Perspectives in Quantum Hall Effects}, ed. 
by S. Das Sarma and A. Pinczuk (Wiley, NY 1997).

\bibitem{Spielman-00} I.B. Spielman {\it et al.}, 
Phys. Rev. Lett. {\bf 84}, 5808 (2000).

\bibitem{Andrei-93} L. Menna, S. Y\"{u}cel, and E.Y. Andrei,
Phys. Rev.  Lett.  {\bf 70}, 2154 (1993).  E.Y. Andrei, in
Ref. \cite{Andrei_book}, p.~207.

\bibitem{Shklovskii} A strong effect on tunneling of the momentum 
transfer to defects was first discussed by B.I. Shklovskii, JETP
Lett. {\bf 36}, 51 (1982); B.I. Shklovskii and A.L. Efros,
Sov. Phys. JETP {\bf 57}, 470 (1983).

\bibitem{Levitov-96} A physically different problem of tunneling between 
lattice sites of the WCs at the edges of a quantum Hall system was
discussed by M.B. Hastings and L.S. Levitov, Phys. Rev. Lett.  {\bf
77}, 4422 (1996). In contrast to the present case, the tunneling
probability was determined by coupling to low-frequency
long-wavelength modes, it oscillated with $B$, and went to zero for
$T\to 0$.

\bibitem{AP-90} M.Ya. Azbel and P.M. Platzman, 
Phys. Rev. Lett. {\bf 65}, 1376 (1990).

\bibitem{Feynman}
R. P. Feynman and A. R. Hibbs, {\em Quantum Mechanics and Path
Integrals} (McGraw-Hill, New York, 1965).

\bibitem{Leggett} A.O. Caldeira and A.J. Leggett, Ann. Phys. (N.Y.) {\bf 149},
374 (1983).

\bibitem{us} T. Barabash-Sharpee, M.I. Dykman, P.M. Platzman,
Phys. Rev. Lett. {\bf 84}, 2227 (2000)

\bibitem{Langer} J.S. Langer, Ann. Phys. (N.Y.) {\bf 41}, 108 (1967);
S. Coleman, Phys. Rev. D {\bf 15}, 2929 (1977).

\bibitem{Ao} For a parabolic well  and a special form of electron coupling 
to harmonic oscillators, the instanton technique was applied to
tunneling in a magnetic field by P. Ao, Phys. Rev. Lett. {\bf 72},
1898 (1994); Physica Scripta T{\bf 69}, 7 (1997).


\bibitem{Mark} M.I. Dykman, J. Phys. C {\bf 15}, 7397 (1982). 

\bibitem{Iye-80} Y. Iye, {\it et al.}, J. Low Temp. Phys. {\bf 38},  293 (1980).

\bibitem{Eperp} M.J. Lea {\it et al.},  Phys. Rev. B {\bf 55}, 16280 (1997).


\bibitem{Goodkind} For $B=0$, a good agreement between measured and
numerically evaluated tunneling rates for electrons on helium was
obtained by G.F. Saville, J.M. Goodkind, and P.M. Platzman,
Phys. Rev. Lett. {\bf 70}, 1517 (1993).

\bibitem{Eisenstein}J. Smoliner {\it et al.}, Phys. Rev. Lett. {\bf 63}, 
2116 (1989); J.P. Eisenstein {\it et al.}, Phys. Rev B {\bf 44}, 6511
(1991); S.Q. Murphy {\it et al.}, Phys. Rev. B {\bf 52}, 14825
(1995).

\end{thebibliography}
\end{document}